\documentclass[12pt]{article}
\usepackage[superscript]{cite} 
\usepackage{hyperref}
\usepackage{times}
\usepackage[T1]{fontenc} 
\usepackage{breakcites} 
\usepackage{url} 

\newcommand{\itemb}{\begin{itemize}}
\newcommand{\iteme}{\end{itemize}}

\newcommand{\enumb}{\begin{enumerate}}
\newcommand{\enume}{\end{enumerate}}

\newcommand{\smallest}[1]{\vspace{6pt}\noindent {\bf  {#1}.}}

\newcommand{\com}[1]{}
\usepackage{xcolor}

\newcommand{\op}{opioid}
\newcommand{\Op}{Opioid}

\newcommand{\spont}{spontaneous}

\newcommand{\adol}{adolescence}

\newcommand{\antip}{antipsychotic}

\newcommand{\antag}{antagonist}

\newcommand{\xtblty}{excitability}

\newcommand{\desensit}{desensitization}

\newcommand{\metab}{metabolism}

\newcommand{\hthal}{hypothalamus}

\newcommand{\amg}{amygdala}
\newcommand{\Amg}{Amygdala}
\newcommand{\hippo}{hippocampus}

\newcommand{\pitu}{pituitary}

\newcommand{\icell}{{{intracellular}}}

\newcommand{\ecell}{{{extracellular}}}

\newcommand{\sympa}{{{sympathetic}}}

\newcommand{\phos}{{{phosphorylation}}}

\newcommand{\infl}{{{inflammation}}}

\newcommand{\plast}{{{plasticity}}}

\newcommand{\withd}{{{withdrawal}}}

\newcommand{\tname}{{T*MDD}} %

\newenvironment{sciabstract}{\begin{quote} \bf}{\end{quote}}

\com{
\usepackage{fancyhdr}
\pagestyle{fancy}
\lhead{}
\rhead{April 2024}
}
\title{A Dynorphin Theory of Depression \\ and Bipolar Disorder}
\author{
{Ari Rappoport}\\ 
\normalsize{The Hebrew University of Jerusalem, Israel}\\
\normalsize{ari.rappoport@mail.huji.ac.il}
}
\date{\normalsize{April 2024}}

\begin{document}
\maketitle
\begin{sciabstract}
Major depressive disorder (MDD) is a debilitating health condition affecting a substantial part of the world's population. At present, there is no biological theory of MDD, and treatment is partial at best. Here I present a theory of MDD that explains its etiology, symptoms, pathophysiology, and treatment. MDD involves stressful life events that the person does not manage to resolve. In this situation animals normally execute a `disengage' survival response. In MDD, this response is chronically executed, leading to depressed mood and the somatic MDD symptoms. To explain the biological mechanisms involved, I present a novel theory of \op s, where each \op\ mediates one of the basic survival responses. The \op\ mediating `disengage' is dynorphin. The paper presents strong evidence for chronic dynorphin signaling in MDD and for its causal role in the disorder. The theory also explains bipolar disorder, and the mechanisms behind the treatment of both disorders. 
\end{sciabstract}


\section*{Introduction}
Major depressive disorder (MDD) affects hundreds of millions of people worldwide, and is one of the leading causes of years lost to disability, with increasing prevalence over the last decades \cite{james2018glo}. 
The efficacy of current pharmaceutical treatment is moderate at best, with substantial negative side effects  \cite{malhi2018dep, otte2016maj}. 
MDD is closely linked to bipolar disorder (BP), the distinction depending on the presence of manic episodes. 
At present, there is no theory that explains MDD or BP \cite{malhi2018dep, otte2016maj}, despite major research activity (over 700K/2M papers in pubmed/Google scholar). 

Here I present a theory (\tname) that explains the etiology, symptoms, pathology and treatment of MDD and bipolar disorder. The theory is supported by strong preclinical and clinical evidence, and points to novel treatment directions. 

\smallest{Theory overview}
MDD is triggered by stress-inducing life events that the person does not manage to resolve. In such situations, animals execute one of the innate survival responses (`fight or flight'). MDD occurs when the `disengage' (withdraw, give up) response is impaired and chronically executed. 

To explain the biological mechanisms involved, I present a novel theory of \op s and survival responses, in which each \op\ supports one of the survival responses. The \op\ supporting the disengage response is dynorphin (DYN), which acts via the kappa \op\ receptor (KOR). KOR mediates disengage by suppressing the phasic responses of neurons releasing dopamine (DA), corticotropin-releasing hormone (CRH), norepinephrine (NEP), and serotonin (SER). 

In MDD, KOR signaling is impaired, leading to its chronic activation. The resulting dysregulation of DA, CRH, NEP and SER reduces goal-directed behaviors, dysregulates appetite and sleep, and induces motor agitation or retardation, difficulties in concentration, and fatigue. In humans, this state is associated with dysphoria and reduced positive emotions, and promotes suicidal thoughts (the ultimate `withdraw' response). Thus, \tname\ explains the defining symptoms of MDD \cite{DSM5MDD}. 
Females are more vulnerable than males because their expression of CRH1 is higher and their KOR signaling is more prolonged. 

In addition to MDD, \tname\ explains bipolar disorder. Manic episodes occur when chronic KOR activation leads to its complete \desensit, allowing the dysregulated NEP and DA systems to reach a hyperactivated state. This leads to reduced sleep, increased goal-directed activities, excessively elevated mood, a feeling of omnipotence, and other mania symptoms \cite{otte2016maj}. 

The leading treatment of MDD at present is via drugs that increase brain SER levels. Moderate and high SER levels oppose DYN and provide energy, explaining why these drugs suppress most MDD symptoms. However, SER also opposes CRH and NEP, with general suppression of novelty-induced responses, including positive ones. This can result in an unpleasant emotional state. In addition, high SER levels can induce motor and mental agitation. \tname\ points to the KOR pathway as a promising treatment direction.


\section*{Stress Responses and MDD} \label{sec:stress}
Organisms can be in a state of {\bf satiety} with no needs or threats, or in a {\bf stress} state where these are present. Stressful events trigger internal body responses collectively known as {\bf stress responses}, whose goal is to recruit resources (e.g., glucose, oxygen) that allow the organism to deal with the event. 
Stressful events can involve both threats (e.g., predators) and need-satisfying opportunities (e.g., food, sex). 

There are two major stress response systems, the {\bf \sympa\ nervous system (SNS)} and the {\bf \hthal-\pitu-adrenal (HPA) axis}. At stress onset, external surprises trigger an alert conveyed by high frequency neural firing in sensory input paths. This alert recruits the SNS, which provides immediate energy via glycogenolysis and elevated blood glucose and heart rate. It also enhances sensory inputs (e.g., by increasing pupil diameter). In parallel, sensory input induces the release of CRH, mainly from neurons in the central nucleus of the \amg\ (CeA) and the paraventricular \hthal\ (PVH). CRH stimulates the HPA axis, acting on CRH1 receptors in the \pitu\ to induce the release of ACTH, which in turn induces the secretion of glucocorticoids (cortisol in humans) from the adrenal gland \cite{bale2004crf}. Within minutes, cortisol increases blood glucose and promotes food intake. 
CRH also sustains the recruitment of the SNS, and recruits its brain counterpart, NEP from the locus coeruleus (LC). 


MDD is known to be triggered by life events involving strong stress \cite{willner2013neu}. 
In some people, MDD can be triggered by mild stress. However, this type of susceptibility is usually associated with early life or parental stress, and can be attributed to stress-induced epigenetic or genetic adaptations \cite{willner2013neu, bertollo2024ear}. Thus, stress is the underlying cause of MDD in virtually all cases. 

In accordance with chronic stress, MDD patients show chronically dysregulated (mainly elevated) SNS and HPA axis activity \cite{veith1994sym, arborelius1999rol, holsboer2000cor, wong2000pro, gold2005car, rein2019hyp, xie2024neu, cottingham2012alp, cubala2014low, izakova2020sal}. 
However, stress is involved in many, probably most, psychiatric disorders \cite{babenko2015str}. Thus, although depression is triggered by stress, stress responses alone do not explain the unique characteristics of MDD. 

\section*{A Theory of \Op s and Survival Responses} \label{sec:ops}
A central thesis of \tname\ is that the occurrence of MDD is related to the concrete actions that the person takes to deal with stress, rather than to general stress responses. Specifically, \tname\ posits that MDD is due to the chronic execution of the `disengage' survival response. 

A major contribution of the present paper is a theory of \op s (T*OPS) that connects survival responses with \op s (OPs), allowing a mechanistic understanding of MDD. In principle, \tname\ only requires a theory of DYN function. However, presenting the full theory of OPs strengthens the case for both DYN function and its effect in MDD. 

\smallest{Survival responses}
Animals have a repertoire of innate survival responses, the main ones being {\bf charge} (prey or sex), {\bf fight, flight, chase}, and {\bf disengage (withdraw, give up)}. 
These responses promote survival of the individual or the species. Stress responses are body internal, and their main function is metabolic, while survival responses involve external actions. 
Note that both stress and survival responses respond to both rewarding and aversive stimuli \cite{koolhaas2011str}. 

The disengage response is crucial for survival in confrontations with a strong adversary. 
Some animals have a `faint' response, which is a variant of disengage. Another variant is freezing without muscle flexing and with reduced heart rate. Freezing with muscle flexing and increased heart rate accompanies the SNS response to an initial surprise, and is not one of the survival responses strategies. 


\smallest{\Op\ signaling}
Survival responses are crucial to the animal and to its species, so it is reasonable to expect that there would be molecular agents that have evolved to support each specific survival response. 
According to T*OPS, {\bf \op s are the major agents mediating survival responses. }

There are four well-known OPs, beta-endorphin (BEND), enkephalin (ENK), nociceptin (NOCP), and dynorphin (DYN). 
These signal via the G protein-coupled receptors (GPCRs) 
mu \op\ receptor (MOR), delta \op\ receptor (DOR), NOCP receptor (NOCPR, also known as nociceptin/orphanin FQ receptor and opioid receptor like 1, ORL1), and kappa \op\ receptor (KOR), respectively \cite{bian2012opi}. 

Endomorphin1 (EM1) and endomorphin2 (EM2) have received less research attention than the other OPs, but are probably just as important (see below). Like BEND, they are MOR ligands, but may act on different alternatively spliced MOR subtypes \cite{sakurada2008pos, liu2021alt}.

The \op\ receptors (OPRs) are usually described as coupled to Gi/o proteins, with inhibitory actions on their cells via suppression of cAMP production, and activation of G protein-coupled inwardly-rectifying potassium channels (GIRKs) \cite{bian2012opi}. A lesser-known fact is that OPRs also couple to Gs, which has opposite effect on cells, exciting them by decreasing potassium currents and activating adenylate cyclase (AC) isoforms \cite{fan1995dua, szucs2004dua, bian2012opi}. 
Low (nanomolar) and high (micromolar) agonist concentrations activate Gs and Gi/o binding, respectively \cite{fan1995dua, kuzmin2000dos}. 

Once bound to their ligand, OPRs undergo \phos\ (mainly by G protein receptor kinases (GRKs)), which allows binding by beta-arrestin and endocytosis. Endocytosed receptors can be degraded or recycled to the plasma membrane. Importantly, they can also yield prolonged signaling in intracellular compartments (mainly endosomes) \cite{crilly2021com}. 

In light of these data, T*OPS uses the following description of OPR signaling, which is more accurate than labeling it as `inhibitory'. {\bf OPRs suppress phasic cellular responses to novel \ecell\ inputs, while allowing continuation of, and quickly suppressing, responses that have already started.} For the first part of this description, new \ecell\ responses are suppressed by suppressing AC and thus cAMP production and PKA. OPR binding to Gs can activate AC isoforms, but these isoforms are linked to PKC \cite{halls2011reg}, which is associated with the release of calcium from \icell\ stores, less with responses to acute \ecell\ inputs. 
For the second part of this description, OPRs oppose (via Gs) or slow down (via Gi/o) outward potassium currents. This does not rapidly hyperpolarize the cell, but keeps it depolarized (i.e., allows on-going activity). Thus, OPRs disconnect neurons from new acute inputs, but increase their on-going activity and intrinsic \xtblty. Eventually, the response is terminated. 

\smallest{\Op\ release and signaling}
According to T*OPS, each \op\ agent is expressed in cells that act as sensors for the execution of the agent's survival response. When this condition is detected, the cell releases the OP in large quantities.
Release in smaller quantities can occur prior to the execution of the survival response, due to brain \plast. Past events in which the survival response has been executed enhance synaptic connections from neurons activated by the sensory context to the OP releasing neurons. As a result, OP-releasing neurons are excited when the animal is in similar contexts. Thus, past experience teaches the brain to anticipate survival responses in certain situations, which can lead to low-grade OP release before survival response execution. 

The OPR activated by the OP agent is expressed in neurons involved in response execution (we refer to these neurons as the OP's {\bf path}). Small ligand concentrations indicate anticipation of imminent execution, and activate Gs to enhance the path's excitation and accelerate the response.  
Once execution successfully commences, \op\ ligand concentrations are increased, switching the OPR to Gi/o to ensure response completion. This strongly prevents the path from being activated by new inputs, while still allowing its on-going activity. This architecture minimizes interference with the execution of survival responses, which are crucial to the animal and the species. 

This is why all OPs have a strong anti-pain effect. Pain is conveyed by sensory neurons to indicate states that may harm cells, and thus stimulates responses. However, survival responses, by their very definition, enjoy precedence over other responses. Once they start executing, they suppress any interference generated by new inputs. Hence, in T*OPS, pain suppression is part of OP-mediated potentiation of survival responses, not the main physiological role of OPs. 


\smallest{\Op\ function}
According to T*OPS, each \op\ mediates one of the basic survival responses. Specifically, {\bf MOR, DOR, NOCPR, and KOR mediate charge, chase/flight, fight, and disengage, respectively}. As to ligands, BEND may be specific to charge for food (i.e., preying), while EM1 and EM2 may be specific to charge for sex. 

Since the focus of the present paper is MDD, the description of DYN/KOR below is more detailed than that of the other OPs.

\smallest{BEND, endomorphins, MOR}
The role of MOR in T*OPS is to support charge survival responses, for food or for sex. Successful charge for food involves a rapid increase in blood food metabolites. Indeed, POMC neurons in the arcuate nucleus of the \hthal, which are known sensors of food metabolites in blood, release BEND and increase its CSF concentrations during fat eating \cite{mizushige2009pre}. 
MORs in the basolateral \amg\ (BLA) promote vigorous food approach to predictive cues \cite{lichtenberg2017amy}. 
Stimulation of nucleus accumbens (NAc) MORs induces appetite for high-fat food, while their antagonism decreases intake of food items containing fat and sugar \cite{kelley2002opi, shin2010rev}. 

MOR is also clearly involved in sexual behavior. MOR agonists produce sensations similar to orgasm in people \cite{paredes2014ops},  
and orgasm in men is accompanied by significant release of an endogenous MOR ligand at the \hippo\ \cite{jern2023end}. 
MOR can trigger ejaculation without sensory stimulation \cite{kozyrev2015act}, and 
sexual behavior induces MOR-dependent conditioned place preference (CPP) \cite{paredes2014ops}. 
The precise mechanisms through which OP-releasing cells sense successful sexual climax are not known, but include a rapid increase in estrogen \cite{corona2012hor}. 

A major way in which MOR achieves its function is by promoting the action of dopamine. MOR agonists disinhibit DA neurons, and can also directly induce DA release, in both the ventral tegmental area (VTA) and the A11 DA group, which innervates the spinal cord and promotes vigorous movements \cite{fields2015und, pappas2011opi}. 

T*OPS hypothesizes that BEND's main role is charge for food (prey), while endomorphins promote charge for sex. There is some evidence supporting this idea (see also DYN below), but more research is needed to validate it. 



\smallest{ENK, DOR}
The role of ENK and DOR in T*OPS is to support fast locomotion, in the flight or chase (predation) survival responses. 
ENK is widely expressed in non-neuronal tissue, including the heart, skeletal muscle, and glia, and is released in response to exercise and hypoxia \cite{armstead1998nit, denning2008pro}. 
%
DOR is expressed in the spinal ventral horn, where neurons supporting motor movements are located. It is expressed in inhibitory interneurons (IINs) \cite{wang2018fun}, so its activation disinhibits excitatory neurons (motor neurons and interneurons) that drive movement. 
%
DOR agonists promote locomotion and suppress MOR-induced feeding \cite{wei1977com, katsuura2010mod}. 
ENK controls the cortical release of acetylcholine (ACh) \cite{jhamandas1980act}. Since ACh is the agent that drives muscle contraction and locomotion, this shows that ENK supports locomotion. 
ENK specifically enhances horizontal locomotion (running), with no effect of even a negative effect on vertical movements (rearing) \cite{mickley1990bra, mendella1996var, hebb2003cen}. 
In the spinal cord dorsal horn, DOR controls mechanical but not thermal pain, supporting involvement in external movement \cite{basbaum2009cel}. 

It can be argued that although chase and flight responses involve the same motor movements, they are made in completely different survival contexts and should hence involve different OPs. This may be the reason why there are two dominant ENK isoforms. However, their specific support in chase vs flight is not known. ENK is also a MOR agonist, supporting involvement in chase for food.




\smallest{NOCP, NOCPR} In T*OPS, the role of NOCP is to support the fight survival response. 
BNST NOCP neurons are stimulated by aversive odor \cite{ung2022pre}. 
Aversive social confrontations increase NOCPR mRNA in the \amg\ and \hthal\ \cite{green2009noc}, 
while NOCP mRNA is decreased in suicide \cite{lutz2015dec}. 
NOCP agonists increase aggression \cite{silva2019noc}, and induce scratching and biting movements \cite{inoue1999dos, sakurada1999noc, duvan2017eff}. 
NOCP achieves its function at least partly via substance P \cite{inoue1999dos, sakurada1999noc}, which is known to be involved in aggression \cite{halasz2009sub}. 

\smallest{DYN, KOR}
DYN is mainly expressed in the central \amg\ (CeA), \hthal, prefrontal cortex (PFC), basal ganglia (striatum), \hippo, and the spinal cord dorsal horn \cite{tejeda2012the}. 
It is strongly expressed in neurons whose activity indicates acute needs and effort, including CeA CRH neurons \cite{reyes2011amg}, 
\hthal\ and \pitu\ arginine vasopressin (AVP) and oxytocin (OXT) neurons \cite{dumont1996opi, iremonger2009ret, russell2003mag}, 
\hthal\ orexin neurons \cite{muschamp2014hyp}, 
and striatum principal neurons expressing the dopamine D1 receptor \cite{tejeda2012the}. 
Its release is stimulated by sustained intense neural activity \cite{drake1994dyn}, indicating that prolonged efforts have already been made to address the external challenge and maybe it is time to give up. 
Importantly, it has been explicitly shown that DYN (the A isoform) yields persistent signaling (AC suppression) when most KORs are in late endosomes and lysosomes \cite{kunselman2021com}.  

Unlike MOR and DOR, which are mainly expressed in IINs (except in the spinal cord), KOR has substantial expression in principal neurons, including excitatory projections from CeA to the locus coeruleus \cite{kreibich2008pre, reyes2011amg}, 
\amg, thalamus, and contralateral cortical projections to mPFC \cite{yarur2023dyn},
and BLA projections to the BNST \cite{crowley2016dyn}
and the striatum \cite{tejeda2017pat}. 
%
KOR exerts an especially strong control over CRH \cite{hein2021kap}, 
VTA DA \cite{margolis2006kap, chefer2013kap}, 
and brain NEP \cite{kreibich2008pre} neurons,   
and regulates dorsal raphe SER neurons \cite{lemos2012rep, tao2005mu}.
It also controls adrenal medulla epinephrine release \cite{aunis1999phy}, SNS responses \cite{fassini2015kap}, and the HPA axis \cite{varastehmoradi2023kap}. 
%



\Amg\ DYN neurons are activated during fear memory \cite{jungling2015inc}, supporting the T*OPS model. 
%
DYN is released in virtually all animal models of chronic or repeated stress, including social defeat stress, forced swim test, and restraint stress. In these models, the effect of KOR is clearly that of `give up', manifested as immobility, defeated posture, and decreased social interaction, heart rate, and blood pressure \cite{mclaughlin2006soc, carlezon2006dep, fassini2015kap, donahue2015eff, laman2017lon, williams2018acu, browne2018rev, zan2022amy}. 
The aversive effects of KOR are learned and remain present after the event, as shown by KOR promotion of conditioned place aversion (CPA) \cite{ghozland2002mot, land2008dys, robles2014eff, laman2017lon, hein2021kap}. 
KOR effects on midbrain DA are central in its aversive effects \cite{chefer2013kap, robble2020lea}, including social avoidance \cite{resendez2016dop}. 
KOR reduces intracranial self-stimulation, usually interpreted as reduced motivation, goal-directed behavior, and pleasure \cite{todtenkopf2004eff, carlezon2006dep, negus2010eff}. 
%
These effects are generally blocked by KOR antagonist given before the event \cite{williams2018acu, browne2018rev} (see \cite{browne2021kap} for a highly detailed review). 

It is well-known that infralimbic mPFC (ILmPFC) mediates fear extinction, mainly via \amg\ projections \cite{sotres2010pre}. Fear extinction is successful when the animal stops executing survival responses that involve movements, learning to disengage instead. ILmPFC KORs decrease restraint stress-induced SNS responses \cite{fassini2015kap}, supporting KOR's involvement in extinction. 

DYN use is not limited to negative experience, also participating in positive social survival responses, namely birth and sex. 
DYN regulates AVP and OXT, which control stress-induced fluid management. AVP indicates a need for water, while OXT stimulates fluid secretion during birth, sex, and maternal behavior \cite{stoop2012neu}. 
DYN assists reproduction also by cooperating with MOR. EM2 binds a MOR isoform that induces DYN release \cite{sakurada2008pos}, 
and induces conditioned place aversion \cite{wu2004opp}.
This may be the mechanism behind post-sex aversion to sexual touch. 

There are some reports where low KOR agonist doses induce a positive response (CPP, self-administration) rather than a negative one \cite{braida2008inv}. 
Recall that in T*OPS, DYN suppresses new inputs to DA neurons, but increases their intrinsic \xtblty\ \cite{fields2015und}. Low dose DYN can thus induce DA release and positive DA-mediated responses. 

In summary, the role of \op s is to mediate the basic survival responses.  DYN is released in situations of high stress, and mediates the disengage (withdraw, give up) survival response. Mechanistically, it does this by acting on DA, CRH, NEP, and SER neurons, \amg-PFC connections, the SNS, and the HPA axis. KOR initially excites neurons, and quickly suppresses their phasic responses while increasing their intrinsic \xtblty. 

\section*{A Theory of Depression and Bipolar Disorder} \label{sec:theory}
Now that we understand the role of DYN in survival responses, it is possible to formulate a theory of MDD. The same theory also explains bipolar disorder and some major aspects of drug addiction. 

\smallest{Depression}
MDD is triggered by stress, mainly due to external social challenges. Stress stimulates effortful responses. If these responses do not manage to solve the problem, the person needs to disengage, which is normally mediated by the DYN/KOR pathway. DYN allows people to stop treating stressful challenges as requiring action, and lets them go on with their lives. 

In some people, the DYN path is weaker than in other people, in the following sense. Generally, healthy biological responses are of high amplitude and short duration, and non-healthy ones have a blunted amplitude and a prolonged duration. We refer to such responses as {\bf chronic}. The two aspects of chronicity enhance each other in a positive feedback loop. Specifically with GPCRs, prolonged agonist-induced activation leads to endocytosis and reduced cell surface receptor expression, decreasing the amplitude of future responses. Reduced response amplitude diminishes the probability of resolving the problem, leading to prolonged signaling. In the case of MDD, a weaker DYN path leads to an impaired disengage response. 

In people with a tendency towards chronic DYN signaling, unresolved stress and an impaired disengage response cause persistent thoughts about the problem (rumination), and chronic activation of the systems and agents regulated by DYN, including the SNS, NEP, the HPA axis, DA, SER, AVP and OXT. Since these systems are crucial for recruiting energy and engaging in goal-directed and pleasurable behavior, these capacities are reduced. In addition, since the SNS and HPA axis control sleep and food intake, these are dysregulated. This can manifest as both increased or decreased sleep and appetite, because chronic states involve both increased \xtblty\ and decreased activation amplitude. The subjective experience of this state is what we call `sad' or `depressed'. Since the person keeps viewing the problem as highly stressful (i.e., requiring action), and since they cannot find a conventional way to resolve the situation, they tend to suicidal thoughts and attempts. Thus, chronic DYN signaling and the resulting impaired disengage response can explain all of the symptoms that define MDD \cite{DSM5MDD}.


Why are some people predisposed to MDD? 
MDD risk is greatly increased by stress during sensitive developmental periods (pregnancy, perinatal, early life, and \adol) \cite{sandman2016neu, short2019ear, eiland2013str}, via epigenetic modifications and/or mutations, which can be heritable \cite{babenko2015str}. 
Predisposition to MDD can involve a very large number of genes. These genes do not have to be directly linked to DYN or KOR, because chronic DYN signaling can result from anything that diminishes the person's capacity of resolving external problems in a reasonable amount of time. Thus, a weakness in any of the pathways described above (CRH, SNS, HPA, DA, SER, the other OPs, etc) can increase MDD risk. This explains why depression is so common. However, it is still exhibited by a minority of humans, because impaired disengage can be seriously harmful to the survival of the species. Animals in which such a predisposition is very common would perish during evolution. 



Women are about twice as likely to have MDD than men \cite{malhi2018dep}. 
This can be explained by higher female sensitivity of stress responses to CRH \cite{curtis2006sex}, 
and by estrogen's stimulation of KOR signaling (at least in \hthal\ sex-related neurons) \cite{mostari2013dyn}. Estrogen increases KOR endocytosis and its \icell\ signaling \cite{abraham2018est}, thereby instigating a more chronic-like KOR profile. 

\smallest{Bipolar disorder (BP)}
BP is characterized by the occurrence of both manic and depressive episodes \cite{DSM5MDD}. Mania involves abnormally increased goal-directed activity, energy, and self-confidence, decreased sleep, distractibility, psychomotor agitation, and abnormal risk taking. All of these symptoms can be explained by excessive NEP and/or DA signaling. In \tname, depressive episodes involve chronic KOR activity, which leads to chronic NEP and DA activity. However, chronic KOR activity can lead to its complete \desensit. When this happens, KOR no longer restrains the phasic activation of NEP and DA. Since KOR mediates one of the fundamental survival responses, it is a major factor in restraining NEP and DA, and the disappearance of this factor allows them to signal strongly and persistently. 

This account holds when mania is specifically accompanied by depressive episodes, and does not purport to explain mania in general. Mania can stem from other causes. For example, mania-like behavior is not uncommon in ADHD \cite{birmaher2013bip}. ADHD has a strong comorbidity with MDD, so it can be argued that all three disorders share a common cause. However, the episodic nature and age of onset of MDD and BP, as opposed to the continuous manifestation and very early onset in ADHD, point to different diagnoses \cite{birmaher2013bip}. 


\smallest{Drug addiction}
Any condition that induces chronic DA, NEP and/or CRH can induce DYN release and trigger MDD, especially in predisposed people. In particular, all addictive drugs increase the release of these agents, and are indeed associated with higher rates of depression \cite{volkow2019neu}. 
Different substances of abuse have different properties, and there are aspects of addiction that are not directly related to MDD (e.g., their rewarding properties and the \plast\ changes that they induce). However, \tname\ can naturally explain two major aspects of addiction. 

First, the dysphoria associated with drugs appears during drug \withd, not during drug use. \tname\ explains this by noting that during \withd, cues associated with drug use excite DA, CRH and NEP neurons, and these release DYN that acts on KORs to yield dysphoria as in MDD. Drug use opposes this process (at least as long as the drug-activated reward system is not desensitized), and this may be a major factor in craving and relapse. 
Second, chronic KOR signaling induces chronic SNS activity, which explains the somatic symptoms of drug \withd\ (sweating, increased heart and blood pressure, pupil dilation, stomach cramps) \cite{solecki2019nor}. 




\section*{Evidence} \label{sec:evidence}
Overwhelming preclinical evidence for the involvement of DYN/KOR in aversion, immobility, decreased social interaction, and dysregulated DA, CRH, and NEP has been presented above. Here we provide evidence from   MDD and bipolar patients that KOR signaling is chronic in MDD, and is associated with symptoms. This includes direct evidence, via DYN release, KOR expression and activation, and the effects of KOR agonists on mood in healthy humans and patients, and indirect evidence, via chronic activation of the systems regulated by DYN. 


\smallest{DYN/KOR data in patients}
A major route through which KOR achieves its MDD effects are \amg\ projections. There is direct evidence for both chronic \amg\ DYN and chronic \amg\ \xtblty. Amygdaloid complex pDYN mRNA is significantly decreased in MDD and bipolar patients (41-68\% and 37-38\% respectively) \cite{hurd2002sub}. 
pDYN mRNA in the \amg\ nucleus of the periamygdaloid cortex is decreased in MDD \cite{anderson2013imp}. 
Using PET rCBF data subtraction, significantly increased left \amg\ (and left dlPFC-mPFC) activity was shown in active MDD patients, with a trend in remitted patients \cite{drevets1992fun}. 
MDD patients have an increased early (300ms) \amg\ response to sad faces, showing higher \xtblty\ of the negative paths \cite{fan2024bra}. The same paper reported a decreased late (600ms) response to happy faces, with increased PFC-\amg\ connectivity. 
One paper using PET reported no \amg\ KOR binding differences, but it did not have sufficient power to detect effects that are not very large \cite{miller2018kap}. 

Another major route of KOR effects is PFC, mainly ILmPFC (subgenual cingulate cortex (SgCC) in humans), which mediates extinction and hence disengage. 
KOR expression in right ACC, dorsomedial PFC, and insula negatively correlates with symptoms in unmedicated patients in a depressive episode, indicating chronic KOR activation in threat and stress areas \cite{smart2021dat}. 

Blood KOR levels have been reported to be both lower \cite{quintanilla2023kap} and higher \cite{alHakeim2020maj} than control levels. Lower levels may indicate lower brain levels, and the higher levels were accompanied by higher serum DYN levels. Both pieces of data support chronic signaling. 
Gene linkage association, expression profiling, and network analysis in MDD patients point to KOR \cite{deo2013lar}. 

There is high comorbidity of MDD and drug addiction, and strong clinical and preclinical evidence link DYN to addiction (reviewed in \cite{tejeda2012the, chavkin2016dyn}). 
For example, a significant association between depression severity and two DYN gene mutations has been found in heroin addicts \cite{kaya2021eff}. 



\smallest{KOR agonists and mood in healthy humans and patients}
KOR agonists readily induce dysphoria in healthy humans \cite{kumor1986hum, pfeiffer1986psy, rimoy1994car}.
They can also cause perceptual distortions, nervousness, anxiety,  depersonalization, confusion, and abnormal thinking \cite{pande1996ana, walsh2001ena, ranganathan2012dos}. 
DYN is the only agent discovered so far that is consistently associated with dysphoria. CRH can induce anxiety, a sense of needing to be prepared for negative future events, but does not induce depression. 

Based on the strong preclinical data, KOR \antag s have been proposed as MDD treatment. Their potential as treatment is not clear (see below), but their beneficial effects in initial clinical trials with MDD patients provide strong support for \tname.
A combination of the KOR \antag\ and partial MOR agonist buprenorphine, and the MOR \antag\ samidorphan (to offset the MOR action of buprenorphine) helped treatment-resistant patients \cite{ehrich2015eva}. 
Buprenorphine was found to provide a small benefit for depressive symptoms in a systematic review and meta-analysis \cite{riblet2023eff}. 
The selective KOR \antag\ aticaprant significantly increased ventral striatum activity during reward anticipation and decreased some anhedonia measures \cite{krystal2020ran}. 
In a phase II RCT, aticaprant (as adjunctive treatment) significantly decreased depressive symptoms, more so in patients with higher anhedonia \cite{schmidt2024eff}. 
Another selective \antag, navicaprant, has also shown benefits in a phase II trial \cite{hampsey2024ati}. 

\smallest{Adenylate cyclase (AC)}
The main effect of prolonged Gi/o signaling is suppression of cellular AC. Indeed, decreased AC activity and cAMP have been repeatedly shown in MDD and bipolar patients, both in brain \cite{young1998inc, fujita2012dow}
and platelets \cite{hines2005pla}. 
Depressed suicide victims show decreased brain AC activity \cite{dwivedi2004pro}
%
and increased Gi2 and decreased Gi3 protein in platelets \cite{garcia1997den}. 
This result may provide direct evidence for increased KOR activity. KOR is expressed in platelets \cite{gruba2022cha}, 
KOR equally binds Gi2 and Gi3 (in transfected CHO cells) \cite{prather1995pro}, 
and the only other significant GPCRs that couple to Gi in platelets are P2Y12, which binds Gi2, and the alpha2 adrenergic receptor, which binds Gz \cite{offermanns2006act}. 

There is also some genetic data pointing to AC. MDD-associated genes include various AC isoforms (as well as CRH1 and CRH2) \cite{fan2020ana}, 
and DNA hypo-methylation of AC9, associated with decreased expression and increased CRH, has been found in suicide victims \cite{romero2021int}. Importantly, AC9 signals from endosomes \cite{lazar2020gpr}, supporting sustained \icell\ signaling of a Gi/o-bound GPCR. 

Relatedly, BDNF, which cooperates with cAMP in early \plast, is also decreased in MDD, in brain and serum \cite{castren2017bra}, and is highlighted by genetic analysis \cite{deo2013lar}. 

\smallest{Prefrontal cortex}
KOR has a strong effect in PFC, especially SgCC. Increased functional connectivity between the SgCC and the default-mode network (the brain network engaged in thinking when not doing anything special) was reported in patients \cite{greicius2007res}. Moreover, there was a positive correlation of the effect with the length of the current depressive episode. 
Increased functional connectivity between the SgCC and the insula and \amg\ was reported in depressed adolescents \cite{connolly2013res}. 
These results imply chronic SgCC activity (including with the \amg) that does not manage to mediate disengage. 
Using data subtraction with PET regional cerebral blood flow, increased activity was shown in left mPFC and \amg\ in active patients, with a trend in patients in remission \cite{drevets1992fun}. 
MDD and bipolar patients showed increased amplitude of low-freq fluctuations in resting state fMRI, mainly in the insula and mPFC, indicating higher \spont\ activity \cite{gong2020com}. 
Remitted MDD patients showed decreased PFC responses to both pleasant and aversive stimuli \cite{mccabe2009neu}.
A highly thorough review of fMRI and PET imaging in MDD concluded that patients show increased activity and decreased volume in ventral PFC and \hippo\ \cite{savitz2009bip}. This review also found hypo\metab\ in dorsal PFC, indicating reduced actions. 

\smallest{SNS, brain norepinephrine}
As mentioned above, MDD patients show chronic dysregulation of the SNS \cite{veith1994sym, gold2005car, meyer2006ele, cottingham2012alp, cubala2014low, xie2024neu}
and brain NEP \cite{wong2000pro, gold2005car, cottingham2012alp, zhao2015dif, rein2019hyp}, 
mostly involving higher \xtblty.  


\smallest{HPA axis, cortisol}
MDD patients show higher CRH release, CSF CRH, blunted stimulation of ACTH by CRH, higher plasma cortisol and ACTH, lower morning cortisol, and decreased GR receptor (the cortisol receptor) function \cite{arborelius1999rol, holsboer2000cor, wong2000pro, gold2005car, zhao2015dif, rein2019hyp, izakova2020sal}. 
%
These data clearly point to chronically higher HPA axis activity, with desensitized function. 

\smallest{Dopamine, reward}
DA supports vigorous goal-directed movements and reward motivation \cite{klein2019dop}. 
DA function can be assessed via an animal's DA and/or behavioral response to positive and negative stimuli. There is strong evidence for decreased phasic DA release and reward-related learning signals in MDD \cite{mccabe2009neu, stoy2012hyp, geugies2019imp}. 
A recent systematic review and meta-analysis found higher DA release in the basal ganglia without synthesis capacity differences, indicating chronic activity \cite{mizuno2023dop}. 


%
%

\smallest{Serotonin}
The role of SER in MDD seems paradoxical. 
On one hand, drugs that elevate brain SER levels oppose many MDD symptoms, at least in the short term \cite{hieronymus2016con} (but see \cite{jakobsen2020sho}). 
On the other hand, despite many research efforts, and reports of dysregulation of various components of the SER system, no serious impairment in brain SER has been consistently identified \cite{moncrieff2023ser}. 
Here I explain why this indirectly supports \tname. 

In \tname, SER-based drugs help because medium SER levels generally oppose DYN, NEP, and CRH \cite{mijnster1998ser, dAddario2007rol, guiard2008fun, tassin2008unc, zhang2004ser, flandreau2013esc}. 
Moreover, since DYN regulates SER, we can expect SER signaling to be impaired in MDD. However, DYN's regulation of SER is weaker than its regulation of other agents (see below), so the impairment is not large. In addition, since chronic effects involve both increased tonic release and decreased signaling, the effects of a small-scale chronic impairment are difficult to identify. 
Moreover, SER is much more abundant than DYN, so the effect of DYN should be limited to specific neural paths. 
Thus, the profile of the reported state of SER in MDD fits a DYN-induced problem. 

To better understand the mutual effects of SER, DYN, and other agents, we need to examine the physiological role of SER, which has been the subject of much debate \cite{carhart2017ser}. 
In my view, the main relevant observation is that the vast majority of body SER is present in the gut, where SER release is stimulated by excessive glucose \cite{babic2012glu}. 
In addition, release is stimulated by high temperature \cite{voronova20215}. 
In turn, SER signaling promotes glucose uptake,
and suppresses thermogenesis \cite{yabut2019eme}. 
In other words, SER signaling is activated in situations of excess, and its role is excess disposal. 
This stands in sharp contrast to the stress systems relevant to MDD, which are activated in situations of deficiency (glucose, oxygen, cold). 
Indeed, SER suppresses fight and flight survival responses, thereby promoting `do nothing' (i.e., healthy disengage, covering for impaired DYN signaling) \cite{paul2013fun}. 
In fact, DYN has been reported to decrease the SER transporter \cite{kivell2014sal}, implying that DYN may use SER's `do nothing' function as part of DYN's disengage. 
Thus, SER has a fundamental opposition to deficiency-induced stress, explaining why SER drugs are beneficial in MDD. However, deficiency-induced stress is more common in nature than excess-induced stress, and DYN may use SER, explaining why the SER system is only weakly regulated by DYN and why only weak SER impairment (if at all) is identified MDD. 









\smallest{Oxytocin}
OXT is strongly regulated by KOR. As with other such agents, OXT levels are mainly reported to be decreased in MDD, with some reports of increased levels with high variability \cite{mcquaid2014mak, vanLonden1997pla}. 
One paper reported higher plasma AVP in MDD, probably reflecting its recruitment by stress responses \cite{vanLonden1997pla}. 



\smallest{Immunity}
Cytokines and \infl\ are leading recurrent themes in MDD theories \cite{miller2009inf, malhi2018dep, alawie2021rec}. 
Chronic \infl\ should be expected in MDD, since chronic stress yields chronic immune activation \cite{padgett2003str}. 
However, immunity also has a tight connection to DYN, which is synthesized by immune cells (macrophages, T cells, other cells) following stimulation by CRH \cite{gein2014dyn}. 
Indeed, unmedicated MDD patients show significantly higher serum DYN and KOR than controls, and these are strongly associated with increased levels of the cytokines IL-6 and IL-10 \cite{alHakeim2020maj}. 


\section*{Treatment} \label{sec:treatment}
\smallest{Behavioral treatment}
MDD involves an impaired behavioral response, occurring when the disengage survival response is impaired. Thus, the basic treatment for MDD would be to enhance the disengage response via behavioral techniques. Cognitive behavioral therapy techniques such as relaxed breathing normalize SNS activity, and emulate the effect of a successful disengage response. Engaging in actions that are unrelated to the stressful challenge would make it easier for the brain to disengage, while talking or thinking about the challenge, which increase chronic DYN signaling, are generally not recommended unless they can lead to concrete ways of solving the problem. 

\smallest{Serotonin}
As discussed above, \tname\ explains the 
beneficial effects of SER-based drugs on MDD symptoms via SER's inherent opposition to deficiency-induced stress responses. These beneficial effects come with a painful price, however. SER is associated with a state of satiety, so its suppresses goal-directed actions (DA) and energy recruitment (NEP, SNS). This is fine for a few hours after a meal, but chronically elevated SER continuously suppresses motivated responses to all challenges, including challenges of a positive nature. Thus, it is not surprising that SSRIs reduce excitement and sex drive and are associated with apathy, emotional blunting, and a `zombie-like' feeling \cite{masdrakis2023apa}. 
In addition, to compensate for its opposition to NEP, high doses of SER excite muscles and the SNS, yielding SER syndrome symptoms such as motor agitation and mental confusion \cite{buckley2014serot}. 


It takes weeks for SSRIs to start producing an effect. This is because initially, SSRIs induce the activation of SER1 receptors, whose affinity to SER is high and which act as autoreceptors on SER axons to reduce SER release \cite{blier1998rol}. 
After weeks of SSRI treatment, SER1a receptors are desensitized, and SER levels begin to increase. 
During the few initial weeks, SER levels are actually decreased \cite{blier1998rol}, which might explain the higher risk of suicide at this time period \cite{forsman2019sel}. 
Indeed, SER depletion significantly decreases pDYN mRNA in several brain areas, indicating increased DYN usage \cite{dAddario2007rol}. 
In many cases, SSRIs stop suppressing MDD symptoms after a few months. This is probably due to \desensit\ of the lower affinity SER2 receptors \cite{sanders1989ser}. 


A highly controversial aspect of antidepressants is their long-term effect. There are reports that usage of antidepressants (not only SSRIs) is associated with worse long-term prognosis \cite{hengartner2018ant}, leading to a `chronification' of the disorder and higher risk of switching to bipolar disorder \cite{fava2020may}. 
This may be due to a long-term \desensit\ of the SER system. In addition, because GPCR \antag s increase receptor expression \cite{unterwald2011upr}, chronic SSRI usage probably induces supersensitivity of the stress systems suppressed by SER, making the person more vulnerable to future stressful events. 

\smallest{Mania}
Lithium is highly effective for preventing and treating manic episodes \cite{grande2016bip}, 
yet it is not clear how it achieves this effect. The parsimonious \tname\ account is that lithium suppresses adenylate cyclase, even at very low doses\cite{forn1971eff}. This effectively opposes NEP (and CRH1) signaling, and thus mania. 

Quetiapine is effective in treating BP, both mania and depression \cite{grande2016bip}. 
It is usually described as an atypical \antip\ drug, a drug class associated with SER receptor antagonism. However, quetiapine is also a NEP receptor \antag. In fact, its affinity to the alpha1 adrenergic receptor is almost 4 times higher than to the SER2a receptor \cite{nemeroff2002que}. 
Thus, as for lithium, the \tname\ assumption is that its efficacy in BP stems from its effect on the NEP system. 

\smallest{KOR antagonism}
As mentioned above, selective KOR \antag s have shown promising results in initial RCTs. Such drugs relieve chronic KOR signaling, which should be beneficial in MDD. However, these drugs also weaken acute KOR signaling by reducing response amplitude. Since strong KOR signaling is crucial for mediating disengage, and since impaired disengage is the fundamental issue in MDD, KOR \antag s may also have a detrimental effect. Moreover, strong KOR blockade might induce mania. 
Nonetheless, in the longer term, GPCR \antag s increase cell surface receptor expression, which should improve resilience to future stress. Thus, it seems that KOR \antag s would be best used during depressive episodes and for a short time after episode termination, but not later. 
Naturally, the optimal usage guidelines would depend on the precise characteristics of each drug. 


\smallest{GRK3 inhibition}
An interesting direction is to inhibit GRK3, which is the kinase promoting KOR (and CRH1) endocytosis and \icell\ signaling. This would preserve response amplitude while decreasing its chronic activation. The design of GRK inhibitors (especially brain-penetrating ones) is challenging, but they might work better than KOR \antag s. 

\smallest{NEP-based drugs}
Along with SSRIs, NEP reuptake inhibitors that are also SER or DA reuptake inhibitors are used as first line of treatment for MDD \cite{malhi2018dep}. 
Considering that NEP and DA have a more direct effect on MDD symptoms than SER, we can expect NEP- and DA-based drugs to be effective. Indeed, a recent systematic review and meta-analysis found that stimulants help depressive and fatigue symptoms, and methylphenidate (ritalin) also decreased symptom severity and increased response rate \cite{bahji2021com}. 

\section*{Discussion} 

This paper presented a complete theory of depression and bipolar disorder. The theory is complete in the sense that it explains the etiology, symptoms, pathology and treatment of the disorders, and addresses all of the salient existing evidence. I am not aware of any other complete theory of MDD or BP. 


Besides explaining the desired phenomena and addressing the salient existing evidence, a good biological theory needs to show biological logic, a property that's hard to define but easy to recognize. Biological systems are extremely complicated, involving a very large number of components. A good theory positions itself inside existing bodies of biological knowledge that are related to the explained phenomenon. In the current case, \tname\ makes use of a novel complete theory of \op s. Strictly speaking, \tname\ does not need to explain the roles of \op s besides DYN. However, the presentation of T*OPS strengthens the case for the fundamental role of DYN in stress, its mediation of the disengage survival response, and the precise mechanisms of its signaling. Thus, T*OPS provides a strong biological logic for \tname. 

\tname\ includes a complete theory of emotions, which would have positioned it even more firmly in the wider biological context. I have opted to discuss emotions elsewhere, because the subject requires much more space than is available here, and is contentious and would unnecessarily complicate the presentation. 

The main current theories of MDD involve monoamines (SER, NEP, DA), the HPA stress axis, and cytokines (\infl) \cite{malhi2018dep, alawie2021rec}. 
None of these directions can explain the defining symptoms of MDD or the salient evidence. Given that people have already taken active steps for treating MDD using KOR \antag s, an interesting question is why a DYN-based theory has not been formulated already. This may be due to the dominance of the serotonin direction, and to the lack of a theory of \op s that positions DYN in the wider biological context. In any case, this paper makes the case that \tname\ should be the default theory of MDD until some other better theory emerges, or until the evidence supporting it is so large that it can be viewed as correct. 

\tname\ also explains some of the central aspects of drug addiction. Since other central aspects are basically understood, the distance to a complete theory of drug addiction is in my opinion very small. This will be detailed elsewhere. 

The evidence supporting \tname\ is overall strong, but more evidence pertaining to the behavior of DYN and KOR in MDD patients is called for. Compared to the overwhelming preclinical evidence about DYN and KOR, and to the indirect evidence supporting DYN involvement in MDD, the direct evidence is smaller. 

As discussed in the treatment section, KOR \antag s constitute a promising treatment direction, but their success is far from assured. GRK3 inhibitors may have higher chances of success, but are more difficult to design. Discussing this direction in depth is beyond the scope of this paper. 
As mentioned above, platelet Gi3 might be a biomarker, but this need to be verified. 

Finally, \tname\ makes a very strong theory prediction: that normalizing DYN secretion and KOR signaling to be strong and non-chronic would completely resolve the mood symptoms of MDD, and would improve its somatic symptoms. This would dramatically improve the quality of life of hundreds of millions of people. 

\section*{List of Abbreviations}
AC: adenylate cyclase. 
ACh: acetylcholine. 
ACTH: adrenocorticotropic hormone. 
ADHD: attention deficit hyperactivity disorder. 
BEND: beta-endorphin. 
BLA: basolateral \amg. 
BNST: bed nucleus of the stria terminalis. 
BP: bipolar disorder. 
cAMP: cyclic adenosine monophosphate. 
CeA: central nucleus of the \amg. 
CRH (CRH1): corticotropin-releasing hormone (receptor type 1).
CPP: conditioned place preference. 
CSF: cerebrospinal fluid. 
DA: dopamine. 
DOR: delta \op\ receptor. 
DYN: dynorphin. 
ENK: enkephalin. 
Gi/o: a G protein alpha subunit. 
GIRK: G protein-coupled inwardly rectifying potassium channel. 
GPCR: G protein-coupled receptor. 
Gs: a G protein alpha subunit that stimulates PKA. 
HPA: \hthal\ \pitu\ adrenal (axis). 
IIN: inhibitory interneuron. 
ILmPFC: infralimbic medial prefrontal cortex. 
KOR: kappa \op\ receptor. 
LC: locus coeruleus. 
MDD: major depressive disorder. 
MOR: mu \op\ receptor. 
mPFC: medial prefrontal cortex. 
NAc: nucleus accumbens. 
NEP: norepinephrine. 
NOCP (NOCPR): nociceptin \op\ (receptor). 
OP (OPR): \op\ (receptor). 
PFC: prefrontal cortex. 
PKA: protein kinase A. 
PKC: protein kinase C. 
POMC: pro-opiomelanocortin. 
RCT: randomized controlled trial. 
SER: serotonin. 
SgCC: subgenual cingulate cortex. 
SNS: \sympa\ nervous system. 
SSRI: selective serotonin reuptake inhibitor. 
T*MDD: the theory of depression presented in this paper. 
T*OPS: the theory of \op s presented in this paper. 
VTA: ventral tegmental area. 
\indent

\section*{Competing Interests}

The author declares no competing interests.

\bibliography{mdd,asd-anrx-etc,sz}

\bibliographystyle{vancouver} 

\end{document}